\documentclass[prd,nofootinbib,twocolumn,showpacs]{revtex4}
\usepackage{graphics}
\usepackage{bm}

\begin{document}

\title{
Triplication of SU(5) monopoles
}

\author{Levon Pogosian}
\affiliation{Theoretical Physics, The Blackett Laboratory,
Imperial College, Prince Consort Road, London SW7 2BZ, United Kingdom.}

\author{Dani\`ele A.~Steer}
\affiliation{ Laboratoire de Physique Th\'eorique, B\^at.~210,
Universit\'e Paris XI, 91405 Orsay Cedex, France. }

\author{Tanmay Vachaspati}
\affiliation{Department of Physics, Case Western Reserve University,
10900 Euclid Avenue, Cleveland, OH 44106-7079, USA.}

\begin{abstract}
We investigate all spherically symmetric fundamental monopole
solutions with fixed topological charge in the
${\tilde {SU}}(5)\rightarrow [ {\tilde {SU}}(3)\times
{\tilde {SU}}(2)\times {\tilde U}(1) ]/{\tilde Z}_3\times {\tilde Z}_2$
symmetry breaking. We find that
there is three-fold replication of the monopoles. The three
copies correspond to bound states with ``monopole clouds''
arising from the non-Abelian nature of the ${\tilde {SU}}(3)$ factor.
The triplication of monopoles could help us understand the observed
family structure of standard model particles.
\end{abstract}


\

\maketitle

Magnetic monopoles are of diverse and fundamental interest
since they are predicted by all grand unified theories and
embody a rich mathematical structure. They are also central
to discussions of non-perturbative field theory and may
provide a new perspective on particle physics phenomenology.

The topological properties of any field theory leads to a
classification of the topological defect solutions it contains.
However, there are interesting and important properties of
field theories that are missed by the topological classification.
{}For example, a more detailed study of domain walls in an
$SU(5)\times Z_2$ model reveals three different domain wall
solutions, all having the same topological charge
\cite{Vac01,PogVac01}. Guided by the domain wall example, we turn
our attention to magnetic monopole solutions in
\begin{equation}
{\tilde {SU}}(5)\rightarrow [ {\tilde {SU}}(3)\times
{\tilde {SU}}(2)\times {\tilde U}(1) ]/{\tilde Z}_3\times {\tilde Z}_2
\label{symbreak}
\end{equation}
(The tildes indicate that the groups should not be confused with
the symmetry groups of the standard model, which will be written
without tildes.) This symmetry breaking is of interest because it
is the basis of current grand unification models. More importantly
for us is the fact that the topological charges of (non-BPS)
magnetic monopole solutions in this model are in one to one
correspondence with the electric charges of the standard model
leptons and quarks \cite{Vac96}. If duality holds, then any
structure of the ${\tilde {SU}}(5)$ monopole solutions should be
reflected in the standard model particles and vice versa. In
particular we ask: can the family structure of standard model
particles be present in the monopole solutions? (A different
approach to the family structure was taken in \cite{LiuStaVac97}.)

Our approach to solving the problem is to construct all
spherically symmetric monopole solutions with unit winding and
fixed charge orientation in the model. Here we are not interested
in continuous degeneracies of the solutions that are given by the
zero modes. Instead we are looking for features in the space of
solutions, particularly for non-BPS monopoles, that might point to
a replication in the quantum theory of monopoles. In our effort to
determine all such solutions, we rely on the construction of
Wilkinson and Goldhaber (WG) \cite{WilGol77}, though with a novel
twist that we detail below. We are led to three possible solutions
that are then found numerically. Two of these have the strange
feature that they are not localized and hence need careful
interpretation. Such a situation was analyzed in detail by
Weinberg and collaborators \cite{Wei82, LeeWeiYix96} in the BPS
case and, following their discussion, we interpret each of our two
non-localized solutions as a ``monopole plus cloud''. A further
symmetry breaking of the ${\tilde {SU}}(3)$ group in
eq.~(\ref{symbreak}) to ${\tilde U}(1)_1\times {\tilde U}(1)_2$
would condense the clouds into localized objects. Now the
interpretation of the three monopoles as distinct objects is very
clear since they all have different masses.

We briefly discuss the WG \cite{WilGol77} recipe for constructing
spherically symmetric magnetic monopole solutions. (This scheme
has earlier been applied to the ${\tilde {SU}}(5)$ case in
\cite{DokTom80,Sco80} but in a different context.) We start by
fixing the vacuum expectation value (VEV) of the ${\tilde
{SU}}(5)$ adjoint scalar field, $\Phi$, along the $z-$axis:
\begin{equation}
\Phi (z \rightarrow \infty , {\hat {\bf z}})
 \equiv \Phi_0 = {\eta \over {2\sqrt{15}}}{\rm diag} (2,2,2,-3,-3)
\label{Phizvev}
\end{equation}
where $\eta$ is the magnitude of the VEV.
We also fix the magnetic charge, $Q$, of the magnetic monopole
solution:
\begin{equation}
Q = - {\rm diag} \left ( 0,0,\sigma_3,0 \right ) /2 \ .
\label{Qfix}
\end{equation}
($\sigma_a$, $a=1,2,3$ denote the Pauli spin matrices.)
Then the asymptotic magnetic field along the $z-$axis is:
\begin{equation}
e{\bf B} = {Q \over {r^2}} {\hat {\bf r}} \ . \nonumber
\end{equation}
where $e$ is the gauge coupling. This monopole corresponds to the
left-handed quark doublet in the dual standard model \cite{Vac96}.

We would like to find all spherically symmetric solutions
corresponding to monopoles with charge $Q$ given in
eq.~(\ref{Qfix}). The key idea of the WG recipe is to start by
working in the singular Abelian gauge - the gauge in which
monopoles are point objects. In this gauge $\Phi$ is fixed, $\Phi
= \Phi_0$, and the gauge fields are those of a Dirac monopole. It
is relatively straightforward to write down all possible
point-monopole solutions given the symmetries of the system. WG
then prescribe how to construct a gauge transformation that would
transform each of these point-monopoles into spherically symmetric
finite-energy monopole solutions.

As described by WG \cite{WilGol77}, though changing the sequence
of their reasoning, we must find all $SU(2)$ generators $I_a$
such that
\begin{equation}
[I_a , Q]=0 \ , \ \ [I_a , \Phi_0] =0 \ , \ \ a=1,2,3.
\label{IaPhi0}
\end{equation}
The choice of $I_a$ fixes the third generator, $T_3$,
of the $SU(2)$ that defines the spherical symmetry of the solution:
\begin{equation}
T_3 = I_3 - Q.
\label{T3I3Q}
\end{equation}
Note that $[Q, \Phi_0 ]=0$ and therefore $[T_3 , \Phi_0 ] =0$.
Once $T_3$ is known, we can find the ladder matrices $T_\pm$
satisfying:
\begin{equation}
[T_3, T_{\pm} ] = \pm T_\pm \ ,
\label{T3Tpm}
\end{equation}
and hence the $T_1$ and $T_2$ generators via $T_1 = (T_+ + T_-)/2$
and $T_2 = (T_+-T_-)/2i$. Note that if $T_3$ has degenerate
eigenvalues, there can be several choices for the
$T_\pm$. The asymptotic scalar fields are written as:
\begin{equation}
\Phi ({\hat {\bf r}} ) = \Omega ({\hat {\bf r}}) \Phi_0
                           \Omega^{-1} ({\hat {\bf r}})
\label{Phihatr}
\end{equation}
where ${\hat {\bf r}} = (\sin\theta \cos\phi , \sin\theta
\sin\phi, \cos\theta )$, and $\Omega ({\hat {\bf r}}) = e^{-i\phi
T_3} e^{-i\theta T_2} e^{+i\phi T_3}$. On the other hand, the
asymptotic gauge fields are given by
\begin{equation}
e {\bf A} = [{\bf I}({\hat {\bf r}}) - {\bf T}]
                          \times {{\hat {\bf r}} \over r}
\label{eA}
\end{equation}
where ${\bf I}({\hat {\bf r}}) =
   \Omega ({\hat {\bf r}}) \omega^{-1} ({\hat {\bf r}}) {\bf I}
     \omega ({\hat {\bf r}}) \Omega^{-1} ({\hat {\bf r}})
$
with
$\omega ({\hat {\bf r}}) = e^{-i\phi I_3} e^{-i\theta I_2}
                          e^{+i\phi I_3} \ .
$ Note that where $I_a$ is written without any argument, it refers
to $I_a ({\hat {\bf z}})$.

Finally, we need to introduce profile functions to get the
radial dependence of the fields \cite{WilGol77}. This leads
to the following ansatz for the scalar field along the $z-$axis:
\begin{eqnarray}
r \Phi (r) &=& r \; {\rm diag}(\Phi_{11},
\Phi_{22},\Phi_{33},\Phi_{44},\Phi_{55}) \nonumber \\
&=& h_1 (r) \lambda_3 + h_2 (r) \lambda_8 +
             h_3 (r) \tau_3 + h_4 (r) Y  \ ,
\label{hidefn}
\end{eqnarray}
where
\begin{eqnarray}
\lambda_3 = {1\over 2}{\rm diag}(\sigma_3,0,0,0) \ , \
\lambda_8 = {1\over {2\sqrt{3}}}{\rm diag}(1,1,-2,0,0) \ , \nonumber \\
\tau_3 = {1\over 2}{\rm diag}(0,0,0,\sigma_3) \ , \ Y = {{1}\over
{2\sqrt{15}}}{\rm diag}(2,2,2,-3,-3) \ . \nonumber
\end{eqnarray}

The magnetic field is given in terms of complex functions, $v_{ij}
(r)$, $i,j=1,...,5$ but only four of them will be non-zero: the
function $v_{lk}$ is non-trivial only if there is a possible
choice of $T_+$ such that $[T_+]_{lk}$ is non-zero
\cite{WilGol77}. Next we define
\begin{equation}
[M_1 (r)]_{ij} = {{v_{ij}+v_{ji}^*}\over 2}  \ , \ \
[M_2 (r)]_{ij} = {{v_{ij}-v_{ji}^*}\over {2i}}
\label{M+ansatz}
\end{equation}
from which the gauge field is:
\begin{equation}
e{\bf A} (r) = [{\bf M}(r) - {\bf T} ] \times
                        {{\hat {\bf r}}\over r} \ .
\label{eAz}
\end{equation}

The energy functional is
\begin{equation}
E = \int d^3 x
    \left [ {\rm Tr} {\bf B}^2 + {\rm Tr} ({\bf D}\Phi )^2
                  + V(\Phi ) \right ]
\label{Efunctional}
\end{equation}
with
\begin{equation}
V(\Phi ) = -m^2 {\rm Tr}(\Phi^2) + h ({\rm Tr}(\Phi^2))^2 +
              \lambda {\rm Tr}(\Phi^4) - {{m^4} \over {4\lambda '}} \ ,
\label{potential}
\end{equation}
where $\lambda ' = h+7\lambda /30$ and
$\eta=m/\sqrt{\lambda '}$ (see Eq.~(\ref{Phizvev})).

Different functions $v_{ij}$ will be non-trivial, depending on the
choice of ${\bf I}$ satisfying eq.~(\ref{IaPhi0}). The only
possibilities for $I_3$ are: $I_3=0$ or $I_3= {\rm diag} (\pm
\sigma_3,0,0,0)/2$. The former choice gives the conventional
monopole \cite{DokTom80,Sco80}, while it turns out that the latter
can lead to new monopoles solutions in addition to the
conventional one. With ${\bf I}={\rm diag} (+ {\bm
\sigma},0,0,0)/2$, the four non-trivial gauge profile functions
are: $v_{12}$, $v_{14}$, $v_{32}$ and $v_{34}$ corresponding to
the 4 different non-zero entries in the two possible constructions
of $T_+$.  We will write the equations of motion only for this
choice of ${\bf I}$.

Inserting the ansatz for the fields into the energy functional and
then extremizing, leads to the equations of motion for all
spherically symmetric monopole solutions:
\begin{eqnarray}
r^2 v_{12}'' &=& v_{12} (v_{12}^2 + v_{14}^2 + v_{32}^2 + e^2
H_{12}^2 - 1) -
v_{14} v_{32} v_{34} \ , \nonumber \\
r^2 v_{14}'' &=& v_{14} (v_{14}^2 + v_{12}^2 + v_{34}^2 + e^2
H_{14}^2 - 1) -
v_{12} v_{32} v_{34} \ , \nonumber \\
r^2 v_{32}'' &=& v_{32} (v_{32}^2 + v_{12}^2 + v_{34}^2 + e^2
H_{23}^2 - 1) -
v_{12} v_{14} v_{34} \ , \nonumber \\
r^2 v_{34}'' &=& v_{34} (v_{34}^2 + v_{14}^2 + v_{32}^2 + e^2
H_{34}^2 - 1) - v_{12} v_{14} v_{32} \ , \nonumber
\end{eqnarray}
\begin{eqnarray}
r^2 h_1''&=& 2 H_{12} v_{12}^2 + H_{14} v_{14}^2 - H_{23} v_{32}^2 + U_1 \ ,  \nonumber \\
r^2 h_2''&=& {1 \over \sqrt{3}} H_{14} v_{14}^2 + \sqrt{3} H_{23}
v_{32}^2
- {2 \over \sqrt{3}}H_{34} v_{34}^2 + U_2 \ ,   \nonumber \\
r^2 h_3''&=& -H_{14} v_{14}^2 - H_{34} v_{34}^2 + U_3 \ , \nonumber \\
r^2 h_4''&=& {\sqrt{15} \over 3} H_{14} v_{14}^2 +
              {\sqrt{15} \over 3} H_{34} v_{34}^2 + U_4 \ , \nonumber
\end{eqnarray}
where primes denote derivatives with respect to $r$, $U_i$ are
derivatives of the potential that we will not explicitly write
out, and $H_{ij}\equiv r(\Phi_{ii}-\Phi_{jj})$. In the above
equations we have already chosen the phases, $\theta_{ij}$, of the
complex functions $v_{ij}$ such that the energy is minimized.
These phases are constant in space and satisfy $\theta_{12}
+\theta_{34}-\theta_{14}-\theta_{32}= \pi$, resulting in the minus
sign in front of the last term in the equations for the
$v_{ij}$'s.

The boundary conditions on the Higgs field at $r=\infty$ are
determined by requiring that $\Phi =\Phi_0$ asymptotically and
those on the gauge field by requiring that
eq.~(\ref{eAz}) goes over into eq.~(\ref{eA}) at infinity. Then,
\begin{eqnarray}
v_{12}(\infty) &=& 1 \ , \ \
v_{14}(\infty) = v_{32}(\infty) = v_{34}(\infty) = 0 \nonumber \\
h_1(\infty) &=& 0 \ , \ \
h_2'(\infty)= h_3'(\infty)=0 \ , \ \  h_4'(\infty) = \eta \ .
\nonumber
\end{eqnarray}
The boundary conditions on the Higgs field at the origin are
determined by requiring non-singular behavior:
\begin{equation}
h_i (0) = 0 \ , \ \ i=1,2,3,4 \ ,
\nonumber
\end{equation}
and the functions $v_{ij}$ at the origin are as yet unspecified.
This is where we encounter the freedom that ultimately leads to
three different monopole solutions. As this is the crucial new
twist in the recipe, we describe it in some detail.

Going back to the WG recipe, notice that if $I_3$ is ${\rm
diag}(\sigma_3,0,0,0)/2$ then $T_3 = {\rm diag}(1,-1,1,-1,0)/2$.
This leads to two possible choices for $T_\pm$. The first is when
$T_+$ is non-trivial in the 1-2 and 3-4 blocks. The second choice
is when $T_+$ is non-trivial in the 1-4 and 3-2 blocks. Requiring
that the gauge field in eq.~(\ref{eAz}) be non-singular at the
origin fixes the values of functions $v_{ij}$ at $r=0$. Then, for
the case where $T_+$ is in the 1-2 and 3-4 blocks we have the
boundary conditions:
\begin{equation}
v_{12}(0)=1=v_{34}(0) \ ; \ \ v_{14}(0)=0=v_{32}(0) \ .
\label{case1v(0)}
\end{equation}
When $T_+$ is chosen in the 1-4 and 3-2 blocks:
\begin{equation}
v_{14}(0)=1=v_{32}(0) \ ; \ \  v_{12}(0)=0=v_{34}(0) \ .
\label{case2v(0)}
\end{equation}
Similarly, we can also choose $I_3 = {\rm diag}(-
\sigma_3,0,0,0)/2$ and then $T_3={\rm diag}(-1,1,1,-1,0)/2$. This
leads to $T_+$ lying either in the 3-1 and 2-4 blocks, or in the
2-1 and 3-4 blocks. The non-trivial gauge profile functions are
now: $v_{21}$, $v_{24}$, $v_{31}$ and $v_{34}$ \cite{WilGol77}.
The second choice for $T_+$ can be shown to lead to a solution
that has scalar and magnetic fields that are identical to those
in the first case (eq.~(\ref{case1v(0)})). The boundary conditions
on the profile functions for the first choice are determined by
requiring that the gauge field remain non-singular at the origin:
\begin{equation}
v_{31}(0)=1=v_{24}(0) \ ; \ \  v_{21}(0)=0=v_{34}(0) \ .
\label{case3v(0)}
\end{equation}
In addition, the boundary conditions at infinity are
$v_{21}(\infty ) =1$ and $v_{ij} (\infty )=0$ for all
$(i,j)\ne (2,1)$. This is the third possible monopole solution.

\begin{figure}
\scalebox{0.420}{\includegraphics{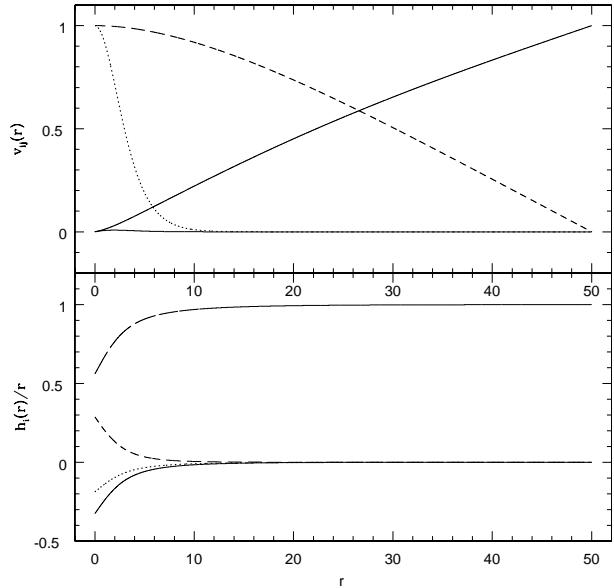}}
\caption{\label{case2profiles} Profile functions $v_{ij}(r)$ and $h_i(r)$
for $e=1$, $\lambda=0.1$, $h/\lambda=-0.2$ and $\eta=1$
(see Eq.~(\protect \ref{potential}) for definitions).
The upper panel shows plots of $v_{12}$ (solid), $v_{14}$ (dot),
$v_{32}$ (short dash) and
$v_{34}$ (long dash). The function $v_{34}$ is not identically zero but
vanishes in the limit of an infinite lattice size. The lower panel shows
plots of $h_{1}/r$ (solid), $h_{2}/r$ (dot), $h_{3}/r$ (short dash) and
$h_{4}/r$ (long dash). Functions $v_{12}$ and $v_{32}$ approach their boundary
values linearly.}
\end{figure}

Now our system is complete -- we have the differential equations
and the boundary conditions. We solve the equations numerically
with the various boundary conditions. The first case (eq.
(\ref{case1v(0)})) leads to the conventional magnetic monopole
which is also discussed in Refs. \cite{DokTom80,Sco80} (The
conventional monopole is usually constructed by setting $I_3=0$
and repeating the above procedure). The second and third set of
boundary conditions give rise to two new solutions. The profile
functions for the second case are shown in Fig.
\ref{case2profiles}. The third monopole solution can be obtained
from the second by switching the 1 and 2 rows and columns of the
fields and simply amounts to re-labeling the profile functions.
Note that, while the conventional monopole obtained using the
boundary conditions given in eq. (\ref{case1v(0)}) is the same as
the one obtained by choosing $I_3=0$, there will be a difference
between the two if one breaks the ${\tilde {SU}}(3)$ symmetry.
This will become relevant later when we consider the
case of ${\tilde {SU}}(3)$ breaking down to ${\tilde
U}(1)_1\times {\tilde U}(1)_2$. With $I_3=0$, the gauge profile
functions must obey the boundary conditions $v_{ij}(\infty)=0$,
$v_{34}(0)=1$ with the rest of $v_{ij}$ vanishing at $r=0$.

The plot in Fig. \ref{case2profiles} shows some unusual features.
While most of the profile functions show ``good'' asymptotic
behavior, the functions $v_{12}$ and $v_{32}$ approach their
boundary values linearly. This behavior can be confirmed by
analytically expanding the equations of motion. It is also
physically reasonable since the functions $H_{12}$ and $H_{23}$
appearing on the right-hand sides of the equations of motion do
not acquire a VEV. Therefore the functions $v_{12}$ and $v_{32}$
correspond to massless gauge degrees of freedom. The energy of the
solution approaches the energy of the conventional monopole as the
lattice gets bigger.

We now interpret this rather peculiar situation following the
discussion of BPS magnetic monopoles in the presence of unbroken
non-Abelian symmetries by Weinberg and collaborators \cite{Wei82,
LeeWeiYix96}. The idea is to consider the case of maximal symmetry
breaking first and then to decrease the VEVs such that, in the
limit of zero VEV, the original non-Abelian symmetry is restored.
In our case, we only need to consider the breaking of the
${\tilde {SU}}(3)$ group further down to
${\tilde U}(1)_1\times {\tilde U}(1)_2$. In
the numerical implementation described above, the symmetry breaking
is achieved by giving a VEV to $h_1$ and setting $v_{ij}(\infty )=0$.
Working, for simplicity, in the BPS limit ({\it i.e.}\ $V(\Phi)=0$), we
now let $h_1 = \eta_1 r$ in the asymptotic region. When $\eta_1 \ne 0$,
the symmetry breaking is:
\begin{equation}
{\tilde {SU}}(5)\rightarrow
[ {\tilde U}(1)_1\times {\tilde U}(1)_2 \times
{\tilde {SU}}(2)\times {\tilde U}(1) ]/{\tilde Z}_3\times {\tilde Z}_2
\nonumber
\end{equation}
As $\eta_1$ is tuned to zero, we obtain the symmetry breaking
pattern in eq.~(\ref{symbreak}).  Now the equations of motion
are solved again with the appropriate boundary conditions.

Just as we might expect, with $\eta_1 \ne 0$, all the three
monopoles are well-localized and have different masses. In the
case of the conventional monopole, obtained using $I_3=0$, the
${\tilde U}(1)_1$ magnetic charge of the solution is zero and the
topological charge of the monopole is
\begin{equation}
Q =  {1\over 2} {\rm diag}(0,0,-1,1,0) \ . \nonumber
\label{Qcase1}
\end{equation}
The topological charge for the monopole in the second case (eq.~
(\ref{case2v(0)})) after the breaking of the ${\tilde {SU}}(3)$
symmetry is
\begin{equation}
Q' = Q + {1\over 2} {\rm diag}(-1,1,0,0,0) \ , \nonumber
\label{Qcase2}
\end{equation}
where the second term is the charge of the ${\tilde U}(1)_1\times
{\tilde U}(1)_2$ monopole which only condenses when ${\tilde
{SU}}(3)$ is broken. The topological charge of the monopole in the
third case is
\begin{equation}
Q'' = Q + {1\over 2} {\rm diag}(1,-1,0,0,0) \ . \nonumber
\label{Qcase3}
\end{equation}
As $\eta_1$ is decreased, the ${\tilde U}(1)_1\times {\tilde
U}(1)_2$ monopole will spread, ultimately filling all space and
becoming a ``cloud''. As in the BPS analysis \cite{LeeWeiYix96},
the energy associated with the cloud goes to zero when we
extrapolate to an infinite lattice. With $\eta_1=0$, we can
explicitly check that monopoles 2 and 3 are gauge equivalent. The
gauge equivalence of monopoles 1 and 2 (and, 1 and 3) is more
delicate since the cloud, which is the cause of gauge
inequivalence on a finite lattice, gets pushed to infinity in the
infinite lattice limit. The two solutions are equivalent but only
under large gauge transformations. As described below, this
delicate behavior is not relevant for us since we will be
interested in the case when ${\tilde {SU}}(3)$ gets broken.

Therefore, we have in hand a total of $3\times (3\times 2) = 18$ gauge
equivalent fundamental monopole solutions in the symmetry breaking
in eq. (\ref{symbreak}). The $3 \times 2$ factor is due to the
3 permutations of the $\lambda_8$ and the 2 permutations of $\tau_3$,
and the factor of 3 refers to the three solutions described above.
Classically there is no distinction between these solutions.
However, in a dual quantum theory the $3\times 2=6$
solutions are believed to correspond \cite{GodNuyOli77} to 3 color
($SU(3)_c$) and 2 isospin ($SU(2)_L$) degrees of freedom \cite{Vac96}.
The remaining factor of 3 suggests that there is a 3-fold replication
of degrees of freedom in the quantum theory even for fixed color,
isospin and hypercharge ($U(1)_Y$).

In the context of the dual standard model, the interpretation of
the 3 solutions is clear. The reason is that color confinement
in the dual standard model is obtained by further breaking
${\tilde {SU}}(3)$ to ${\tilde Z}_3$. Then the colored monopoles
get connected by ${\tilde Z}_3$ strings in color singlet objects.
However, the three solutions are manifestly
distinct when ${\tilde {SU}}(3)$ is broken, since then they have
different masses. Hence they necessarily represent distinct particles
in the dual (electric) theory.

Higher winding monopoles are built by combining fundamental
monopoles and so we expect them to be replicated as well.
It would be very interesting to see if the entire monopole
spectrum is exactly triplicated. If there are indeed three families
of monopole solutions, it would complete the
tantalizing correspondence between the magnetic charge spectrum of
${\tilde {SU}}(5)$ monopoles and the electric charge spectrum of
observed particles \cite{Vac96}.

\

TV is grateful to Fred Goldhaber and Erick Weinberg for useful
discussions. DS is grateful to the Swiss Foundation Ernst et Lucie
Schmidheiny for travel support and to Case Western Reserve University
for hospitality. This collaboration was in part possible due to 
support by the ESF COSLAB network. LP was supported by PPARC
and TV by DOE.

\end{document}